\documentclass[10pt,twocolumn,showpacs, aps, pra]{revtex4-1}
\usepackage[pdftex]{graphicx}
\usepackage{amsmath}    
\usepackage{verbatim}   
\usepackage{color}      
\usepackage{subfigure}  
\usepackage{units}
\begin{document}

\title{Evaporation limited loading of an atom trap}
\author{Markus Falkenau}
\author{Valentin V. Volchkov}
\author{Jahn R\"uhrig}
\author{Hannes Gorniaczyk}
\author{Axel Griesmaier}
\affiliation{5. Physikalisches Institut, Universit\"at Stuttgart, Pfaffenwaldring 57 D-70550 Stuttgart, Germany}

\date{\today}
\begin{abstract}
Recently, we have experimentally demonstrated a continuous loading mechanism for an optical dipole trap from a guided atomic beam \cite{falkenau11}. The observed evolution of the number of atoms and temperature in the trap are consequences of the unusual trap geometry. In the present paper, we develop a model based on a set of rate equations to describe the loading dynamics of such a mechanism. We consider the collision statistics in the non-uniform trap potential that leads to two-dimensional evaporation. The comparison between the resulting computations and experimental data allows to identify the dominant loss process and suggests ways to enhance the achievable steady-state atom number. Concerning subsequent evaporative cooling, we find that the possibility of controlling axial and radial confinement independently allows faster evaporation ramps compared to single beam optical dipole traps.
\end{abstract}
\maketitle

\section{Introduction}
For almost two decades, the standard way of cooling atoms to ultra-cold temperatures has been: laser cooling in magneto-optical traps followed by evaporative cooling in magnetic and optical traps. For many atomic species and especially molecules, different methods are required since laser cooling on a closed cycling transition is technically not feasible \cite{dirosa04, shuman10}. In recent years, there has been considerable progress in the generation of slow and internally cold atomic and molecular beams \cite{meerakker08,Doyle04,Strebel10}, in particular for atoms or molecules with an electric \cite{vanhaecke07,bethlem99,vanBuuren09,patterson09} or magnetic \cite{hogan08,narevicius08} dipole moment. A mechanism that allows for a further increase of the beam's density \cite{ruschhaupt04, szilard29, thorn08, price08, Roos:2003.2, narevicius09, Lahaye:2005.1, anoush09} may hence provide suitable starting conditions for evaporative \cite{ketterle96} or demagnetization cooling \cite{fattori06}. Such a mechanism can be realized in a purely statistical way \cite{Roos:2003.2} or by scattering a single photon \cite{schneble03} that changes the internal state from untrapped to trapped.\par
We have recently demonstrated a loading scheme that is based on the accumulation of chromium atoms from a continuous flux of guided ultra-cold atoms \cite{alex07, axel09, anoush10} in a conservative potential and allows for the fast production of Bose-Einstein condensates (BECs) \cite{falkenau11}. This scheme benefits from the strong magnetic moment of chromium atoms. It is likely to work also for other species with a comparably versatile level structure and strong magnetic moments such as atomic erbium \cite{McClelland.2006} or dysprosium \cite{mingwu10} that have recently gained interest. The scheme might even be the favorable technique of producing BECs from such elements since it omits some of the limitations that are generic to those atoms, e.g. strong dipolar relaxation losses and light assisted collisions. In the case of chromium, the repetition rate has increased from an experimental cycle of more than a minute \cite{Griesmaier.2006b} to 5 seconds for the production of a BEC \cite{falkenau11}.\par
In earlier work \cite{anoush10}, we estimated the transfer efficiency from the guided beam into the trap from the evaluation of single particle trajectories. That approach allowed a prediction of the loading rate and agrees with our measurements. However, we observe a saturation of the atom number at an unexpectedly low level. The estimated maximum number of atoms due to three body collisions of ground state atoms or light assisted collisions should be an order of magnitude higher than the observed ones.\\
In this article, we develop a model based on coupled rate equations to quantitatively describe the loading mechanism. With our model we are able to identify elastic two body collisions, i.e. evaporation as the limiting mechanism in our current trap. The model takes the trap potential's specific non-uniform geometry into account. Regarding evaporation, this implies a dimensionality \cite{Surkov96} of two. Unlike one and three dimensional geometries \cite{Pinkse98}, this has to our knowledge not been treated so far.\par
The paper is organized as follows: In section II we first describe  briefly the experimental procedure used and discussed in detail in \cite{falkenau11, alex07, axel09, anoush10}. We develop a rate equation model for the loading process in section III, and compare numerical simulations based on the model to our experimental data in section IV.
\section{Experimental Setup}
\begin{figure}
	\centering
		\includegraphics[width=0.4\textwidth]{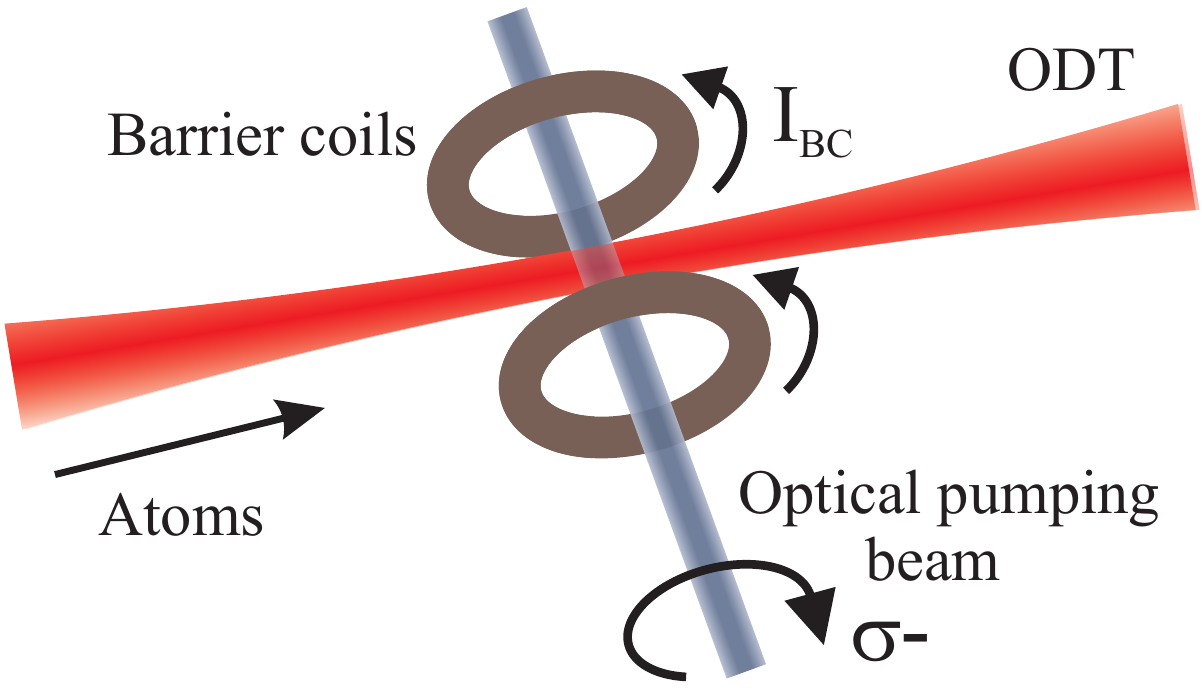}
	\caption{(Color online) Schematic illustration of the setup underlying the continuous loading mechanism. Atoms are traveling to the right and are funneled into the radially confining ODT potential. Axially, the barrier coils create a state dependent potential that is switched from a wall to a well by means of the optical pumping beam.}
  \label{fig:schema}
\end{figure}

In the following, we briefly summarize the underlying mechanism of the loading scheme for convenience \cite{anoush09, falkenau11}. Fig. \ref{fig:schema} shows a schematic illustration of the experimental setup. We overlap a single beam optical dipole trap (ODT) on axis with a continuous beam of $^{52}$Cr atoms. Thereby, about half the flux of atoms in the beam is radially confined in the optical potential. In order to dissipate the atoms' directed kinetic energy, we place a pair of coils around the focus of the ODT to create a magnetic field maximum along the atoms direction of propagation. Since the atoms are initially in a low-field seeking substate, they are slowed as they approach the magnetic barrier potential. Close to their classical turning point, an optical pumping beam transfers the atoms to their absolute ground state, i.e. a high-field seeking state which is attracted axially by the magnetic saddle potential created by the barrier coils and radially confined by the (ODT).
We use the following experimental parameters: the ODT is focused to a waist of \unit[30]{$\mu$m} and operated at a power of \unit[80]{W} which results in a potential depth of about \unit[1]{mK}. Its focus coincides with the geometric center of the magnetic barrier coils. They have a diameter of \unit[1]{mm}, are spaced at a distance of about \unit[1]{mm} and are driven by a current of \unit[1]{A}. The atomic beam has a velocity of \unit[1]{m/s}. All experimental information presented in this article is obtained from in-trap and time of flight absorption images taken along the vertical axis.
\section{Modeling the loading process with rate equations}
\label{sec:loading_rate_eq}
In the following section, we develop a model based on rate equations to explain the physics of the loading process. We will later use the model to interpret our experimental data. Starting from an empty trap, the evolution of the atom number is proportional to the loading rate $L$ until loss processes that depend on the atom number become dominant and eventually an equilibrium state is reached. These considerations, lead to the common rate equation for the evolution of the atom number $N$:
\begin{align}
	\label{eq:ratesimple}
	\dot{N} &= L -\gamma N - \gamma_{ev} N^{2}
\end{align}
where $\gamma$ is the single-body loss rate that includes e.g. collisions with molecules from the background gas and fast impinging atoms from the guided beam. $\gamma_{ev}$ accounts for losses due to binary collisions between trapped atoms. Since the atoms are in their electronic ground state and dipolar relaxation is forbidden due to energy conservation \cite{Hensler.2003}, only elastic two-body collisions can occur in our system. During elastic collisions selectively hot atoms escape which leads to a net cooling of the trapped sample. On the one hand, the last term in Eq. \eqref{eq:ratesimple} thus also changes the temperature, while on the other hand $\gamma_{ev}$ itself strongly depends on the temperature of the trapped gas. The model described by Eq. \eqref{eq:ratesimple} hence needs to be extended to a set of differential equations that also takes into account the evolution of the temperature. In the following section we deduce this model from basic considerations \cite{Yang07, Pinkse98, Berg97, walraven96}.
\subsection{2D-Evaporation}
\label{sec:2DEvaporation}
The physical picture behind evaporation is based on the distinction between trapped and untrapped momentum states \cite{Pinkse98}. A particle initially occupying a trapped state can be promoted to an untrapped one through collisions with other particles. We restrict our considerations to the Knudsen regime (mean free path larger than the size of the cloud), i.e. every particle promoted to an untrapped state escapes. \par
The dimensionality of evaporation is given by the condition of truncation in momentum space \cite{Surkov96}. 3D-evaporation is the simplest case: a particle escapes under the condition $\epsilon > \epsilon_{t}$, i.e. the particle's total energy $\epsilon = U(\vec{r}) + p^{2}/(2m)$ exceeds the trap depth $\epsilon_{t}$, where $U(\vec{r})$ is the potential energy at the position $\vec{r}$ and $p^{2}/(2m)$ the particle's kinetic energy. 3D-evaporation occurs in potentials of uniform depth in all directions (e.g. ideally a single beam ODT or a magnetic trap with rf-knife). \par
The loading mechanism discussed here is based on the superposition of an ODT with a magnetic field maximum \cite{anoush09,falkenau11}. Geometric constraints \footnote{The ODT's Rayleigh length is much larger than the extend of the coils} imply that the hybrid trap potential is to a good approximation separable into an axial (purely magnetic confinement) and a radial part (purely optical confinement):
\begin{align}
\label{eq:SepPot}
	U(\rho,z)=U(\rho)+U(z),
\end{align}
where the potential depth in the radial direction $\epsilon_{\rho}$ is much smaller than along the axial direction $\epsilon_{t,z}$ for typical experimental parameters. Therefore, the criterion of evaporation is more strict: the particle's energy in the radial degrees of freedom must exceed the radial trap depth $U_{\rho}(\rho)+p_{\rho}^{2} > \epsilon_{t,\rho}$. This reduces the probability for hot atoms to be removed and thus evaporation is less efficient.
3D-evaporation, however, can also occur in a non-uniform potential trap due to ergodicity \cite{Hung08}. Whether or not evaporation can be considered ergodic depends on whether a particle finds the trap's exit sufficiently fast in the time between successive collisions. For the trap potential discussed here, numerical evaluation of the classical single particle trajectories shows that the dynamical coupling between the degrees of freedom is negligible.
Surkov \cite{Surkov96} and Pinkse \cite{Pinkse98} compare 1D-evaporation and 3D-evaporation based on an experiment on magnetically trapped atomic hydrogen. For the description of the loading process discussed here, a model considering 2D-evaporation is required.
\subsection{Collision statistics}
\label{sec:collstat}
\begin{figure}
	\centering
		\includegraphics[width=0.45\textwidth]{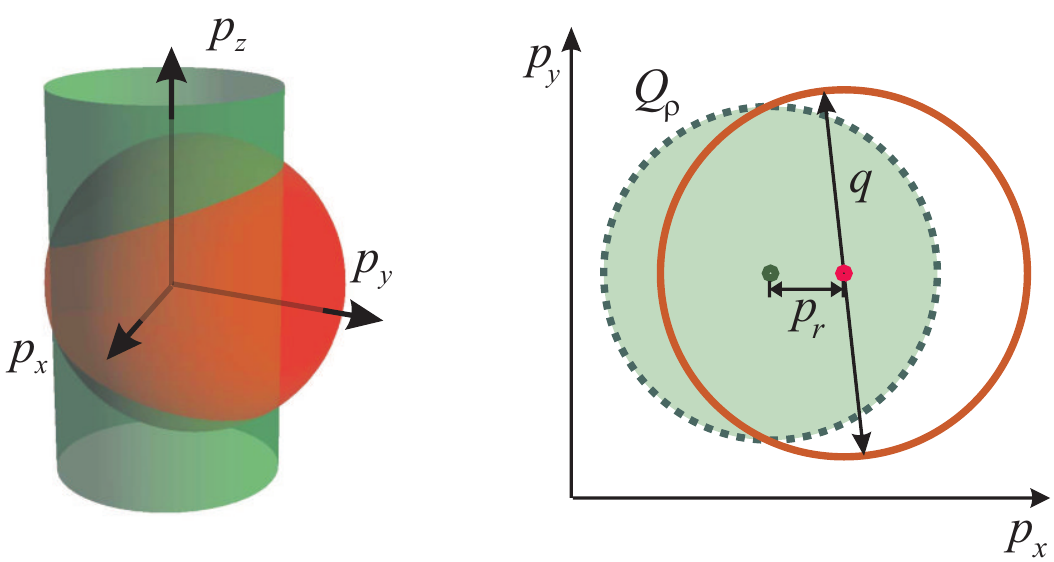}
	\caption{(Color online) (left) The intersection of the surface $S$ of a sphere with radius $q$ and a long cylinder with radius $Q_{\rho}$ geometrically illustrates the loss criterion for particles. The distance between the center of the sphere and the cylinders's axis represents $p_{\rho}$. The fraction of $S$ that lies outside the cylinder corresponds to the statistical probability $f_{\text{loss}}$ of one (or both) atoms to escape the trap. (right) cut along the $p_{x}$ $p_{y}$-plain. Example: The $q$ vector, pointing along the $p_{z}$-axis before collision, has reoriented into the $p_{x}$ $p_{y}$-plain and both particles now occupy an untrapped state. Alternatively, only one atom would be promoted to an untrapped state, if the momenta was reoriented along the $p_{x}$-axis.}
  \label{fig:apfelschale}
\end{figure}
In this section, the Ansatz for 1D-evaporation treated in \cite{Pinkse98} is extended to the case of a non-uniform potential with cylindrical symmetry. The most convenient way to describe the elastic collision of two particles with momenta $\vec{p}_{1}$ and $\vec{p}_{2}$ is to use center of mass coordinates with the average momentum $\vec{p}=(\vec{p}_{1}+\vec{p}_{2})/2$ and relative momentum $\vec{q}=(\vec{p}_{1}-\vec{p}_{2})/2$. Energy conservation requires the relative momentum's absolute value $q=\left|\vec{q}\right|=\sqrt{q_{x}^{2}+q_{y}^{2}+q_{z}^{2}}$ to be conserved. Assuming s-wave collisions, a collision thus reorients $\vec{q}$ isotropically on the surface of a sphere with radius $q$. Furthermore, each component of the average momentum $\vec{p}$ is conserved during collisions.  With the assumption $\epsilon_{t,z} \gg \epsilon_{t,\rho}$ and the cylindrical symmetry of the trap potential (Eq. \eqref{eq:SepPot}), particles leave the trap if their radial momenta $p_{\rho}$ after the collision are larger than the radial escape momentum $Q_{\rho}$.\\
Within the center of mass coordinates, it is merely influenced by two parameters whether a collision redistributes particles into trapped or untrapped states: $q$ and $p_{\rho}$. $q$ denotes the relative momentum of the colliding particles, which is randomly reoriented (azimuthal and polar angles $\phi$, $\theta$) after a collision and $p_{\rho}=\sqrt{p_{x}^{2}+p_{y}^{2}}$, the average radial momentum, respectively. Geometrically, the escape condition thus equals the intersection of a sphere's surface (radius $q$) with an infinitely long cylinder (radius $Q_{\rho}$) with an offset of $p_{\rho}$ as depicted in Fig. \ref{fig:apfelschale}. States outside the cylinder are untrapped. It should be noted that, in contrast to the case of 3D evaporation, an elastic binary collision in a non-uniform potential can also lead to the loss of both particles.\par
Mathematically, Eqn. \eqref{eq:B}-\eqref{eq:f_loss} express the probability $f_{\text{loss}}$ of one (or both) atoms to escape the trap as a function of the input momenta $q$ and $p_{\rho}$ by integration over all possible reorientations of $q$.\\
\begin{widetext}
\begin{align}
  	B(p_{\rho },q,\theta, \phi) \: &= \:\Theta\left[Q_{\rho}^{2}- \left(q \sin\theta \cos\phi - p_{\rho} \right)^{2} - \left(q \sin\theta \sin\phi\right)^{2}\right] \label{eq:B} \\
		f_{\text{loss}}(p_{\rho },q) \: &= \:\frac{1}{\pi}\int_{0}^{\pi} \int_{0}^{\pi} \: B(p_{\rho },q,\theta, \phi) \: \sin\theta \: d\theta \: d\phi \label{eq:f_loss} \\
		\bar{\epsilon}_{\rho}^{+}(p_{\rho },q)  \: &= \: \frac{1}{2 \pi m}\int_{0}^{\pi} \int_{0}^{\pi} \: B(p_{\rho },q,\theta, \phi) \left(q^{2}\sin^{2}\theta-2q p_{\rho}\sin\theta \cos\phi+p_{\rho}^{2}\right)  \sin\theta \: d\theta \: d\phi \label{eq:eps_rho}  \\
		\bar{\epsilon}_{z}^{+}(p_{\rho },q)  \: &= \: \frac{1}{2 \pi m}\int_{0}^{\pi} \int_{0}^{\pi} \: B(p_{\rho },q,\theta, \phi) \left( p_{z}^2 + q^{2}\cos^{2}\theta \right) \sin\theta \: d\theta \: d\phi \;
		\label{eq:eps_z}
\end{align}
\end{widetext}
where $B(p_{\rho},q,\theta, \phi)$ is the criterion for an atom to leave the trap. $\Theta$ represents the Heaviside step function. The mean energy per particle in the radial $\bar{\epsilon}_{\rho}^{+}=\bar{\epsilon}_{x}^{+}+\bar{\epsilon}_{y}^{+}$ and axial $\bar{\epsilon}_{z}^{+}$ degrees of freedom after the collision is obtained by integration over all possible reorientations of momenta as stated by Eqn. \eqref{eq:eps_rho}-\eqref{eq:eps_z}.
\subsection{Thermodynamics}
\label{sec:thermodynamics}
\begin{figure}
	\centering
		\includegraphics[width=0.49\textwidth]{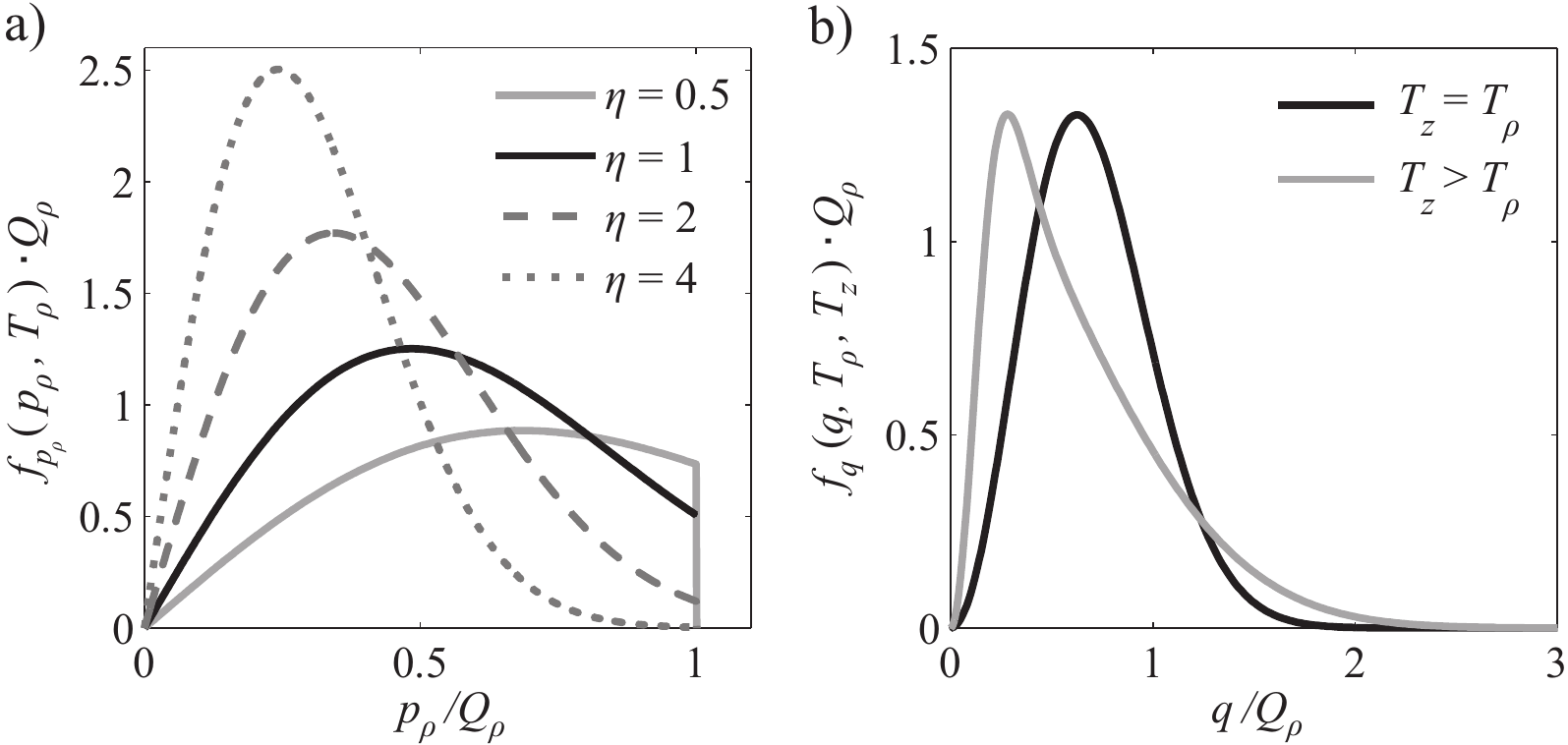}
	\caption{(a) Examples of the statistical distribution of momenta $\left.f_{p_{\rho}}(p_{\rho},T_{\rho})\right|_{T_{\rho}=Q_{\rho}^{2}/2 m k_{B} \eta}$ given by Eq. \eqref{eq:MBpr} for different values of $\eta$. (b) Plots of the statistical distribution of the momenta $f_{q}(q,T_{\rho},T_{z})$ given by Eq. \eqref{eq:MBq}. The black and the gray curves describe particles with the same mean energy $\epsilon$. The black curve shows the thermodynamic equilibrium ($T_{\rho}= T_{z}$), while the gray one represents the case where $T_{\rho} < T_{z}$.}
  \label{fig:t_pr_q}
\end{figure}

Eqns. \eqref{eq:B}-\eqref{eq:eps_z} express the statistical results for general collisions in the trap as functions of the input momenta $q$ and $p_{\rho}$. The goal, however, is to write down $f_{\text{loss}}$, $\bar{\epsilon}_{\rho}^{+}$, and $\bar{\epsilon}_{z}^{+}$ depending on temperature. This is achieved by summing Eqn. \eqref{eq:f_loss}-\eqref{eq:eps_z} over the statistical distribution of $q$ and $p_{\rho}$ at a given temperature. For a thermodynamic description of the collision statistics we make the following assumptions:
\begin{itemize}
	\item The axial and radial degrees of freedom individually are in quasi equilibrium while different temperatures along the respective axes are allowed.
	\item The thermal momentum distribution $f(p_{1,z},T_{z})$ of a single particle along the axial direction is given by a Boltzmann distribution.
	\item The thermal momentum distribution $f(p_{1,\rho},T_{\rho})$ is given by a truncated Boltzmann distribution \cite{walraven96}.
	\item On time average, equipartition of mean kinetic and mean potential energy is assumed (as for the case of a harmonic potential).
\end{itemize}
The first three points are a direct result of the trap's prolate symmetry (regarding trapping frequencies and depth) in combination with the atomic beam. The atoms loaded into the trap have different temperatures in axial and radial directions. These assumptions now allow to describe the atomic cloud with two temperature parameters: $T_{z}$ and $T_{\rho}$. Experimentally, both values can be accessed by time of flight images. The momentum distribution of a single particle within an ideal gas at equilibrium is given by the Boltzmann distribution:
\begin{align}
		f\left(p_{x}, p_{y}, p_{z}\right) = \left(\frac{1}{2 \pi m k_{B}T}\right)^{3/2} e^{-\frac{p_{x}^{2}+p_{y}^{2}+p_{z}^{2}}{2 m k_{B}T}} \ \label{eq:MBf} .
\end{align}
A gas trapped in a potential with a finite depth can be approximated by a Boltzmann distribution truncated at the trap depth $\epsilon_{t}$. It is then convenient to introduce the truncation parameter $\eta=\epsilon_{t}/k_{B}T_{\rho} = Q_{\rho}^{2}/2 m k_{B}T_{\rho}$.
To derive the statistical distributions for $q$ and $p_{\rho}$ at a given temperature, Eq. \eqref{eq:MBf} is transformed into the respective coordinate systems, which leads to (for a derivation see \cite{FalkenauPhD}):
\begin{align}
	f_{p_{\rho}}(p_{\rho},T_{\rho}) \: &= \:n_{0} \: p_{\rho} \: e^{-p_{\rho}^{2}/m k_{B}T_{\rho}} \label{eq:MBpr} \\
	f_{q}(q,T_{\rho},T_{z}) \: &= \: n_{0}' \: q \: \text{erf} \left(q \sqrt{\frac{T_{\rho}-T_{z}}{m k_{B}T_{z} T_{\rho}}}\right) e^{-q^{2}/m k_{B}T_{\rho}}. \label{eq:MBq}
\end{align}
Fig. \ref{fig:t_pr_q} shows examples of these two thermodynamic distributions.
In Fig. \ref{fig:t_pr_q} (a), $f_{p_{\rho}}(p_{\rho},T_{\rho})$ is plotted for different trap parameters $\eta$. The distribution is truncated at the escape momentum $Q_{\rho}$. $f_{q}(q,T_{\rho},T_{z})$ is drawn in Fig. \ref{fig:t_pr_q} (b). The two curves compare a cross-dimensionally thermalized distribution (black) with a non-equilibrium one (gray) at an equal mean energy.
We sum over all possible momentum states by integrating Eq. \eqref{eq:f_loss} multiplied by their statistical weights Eqn. \eqref{eq:MBpr} and \eqref{eq:MBq}, respectively. This yields the desired function for the probability of evaporation \footnote{For a bound interval of integration i.e. a finite radial trap depth, Eqn. \eqref{eq:f_loss_t} - \eqref{eq:eps_z_t}, are not exactly true. The variable $q=\sqrt{q_{x}^2+q_{y}^2+q_{z}^2}$ takes values up to infinity, however, only due to its component $q_{z}^2$ while the radial components are bound to values below $Q_{\rho}$. This fact is neglected in the following.}:
\begin{align}
	f_{\text{loss}}&(T_{\rho},T_{z})\:= \nonumber \\*  =& \: n_{t} \: \int_{0}^{\infty} \int_{0}^{Q_{\rho}} f_{p_{\rho}}(p_{\rho}) \; f_{q}(q) \; f_{\text{loss}}(p_{\rho},q) \ \, dp_{\rho} \,  dq  \label{eq:f_loss_t}
\end{align}
with the re-normalizing pre-factor $n_{t}(T_{\rho})=(\int_{0}^{Q_{\rho}} f_{p_{\rho}}(p_{\rho}) \ dp_{\rho})^{-1} $. In the same way the distributions of the kinetic energies (per particle) among the axial and radial degrees of freedom after a collision are obtained:
\begin{align}	
	\bar{\epsilon}_{\rho}^{+} (T_{\rho},T_{z}) &= n_{t}  \int_{0}^{\infty} \int_{0}^{Q_{\rho}} f_{p_{\rho}}(p_{\rho}) \, f_{q}(q) \; \epsilon_{\rho}^{+}(p_{\rho},q) \, dp_{\rho} \,  dq  \label{eq:eps_r_t} \\
	\bar{\epsilon}_{z}^{+} (T_{\rho},T_{z}) &= n_{t}  \int_{0}^{\infty} \int_{0}^{Q_{\rho}} f_{p_{z}}(p_{z}) \, f_{q}(q) \; \epsilon_{z}^{+}(p_{z},q) \, dp_{z} \,  dq, \label{eq:eps_z_t}
\end{align}
where the integration along $p_{z}$ simply adds a term of $(2-f_{\text{loss}})k_{B} T_{\rho}/2$. Another consequence of the finite radial trap depth is that $\bar{\epsilon}_{\rho} \neq k_{B}T_{\rho}$, i.e. the mean thermal energy $\bar{\epsilon}_{\rho}$ per particle is no longer proportional to the temperature parameter $T_{\rho}$. This can be easily seen from Fig. \ref{fig:t_pr_q}(a). The mean energy is thus a function of the temperature parameter and the escape momentum. Conversion between $\bar{\epsilon}_{\rho}$ and $T_{\rho}$ is done with the following formula and its numerically evaluated inverse:
\begin{align}
	\bar{\epsilon}_{\rho}(T_{\rho},Q_{\rho})=n_{t}(T_{\rho}) \int_{0}^{Q_{\rho}} \frac{p_{\rho}^{2}}{2 m} \: f_{p_{\rho}}(T_{\rho}, \, p_{\rho}) \:d p_{\rho} \label{eq:T2e}
\end{align}
\begin{figure}
	\centering
		\includegraphics[width=0.43\textwidth]{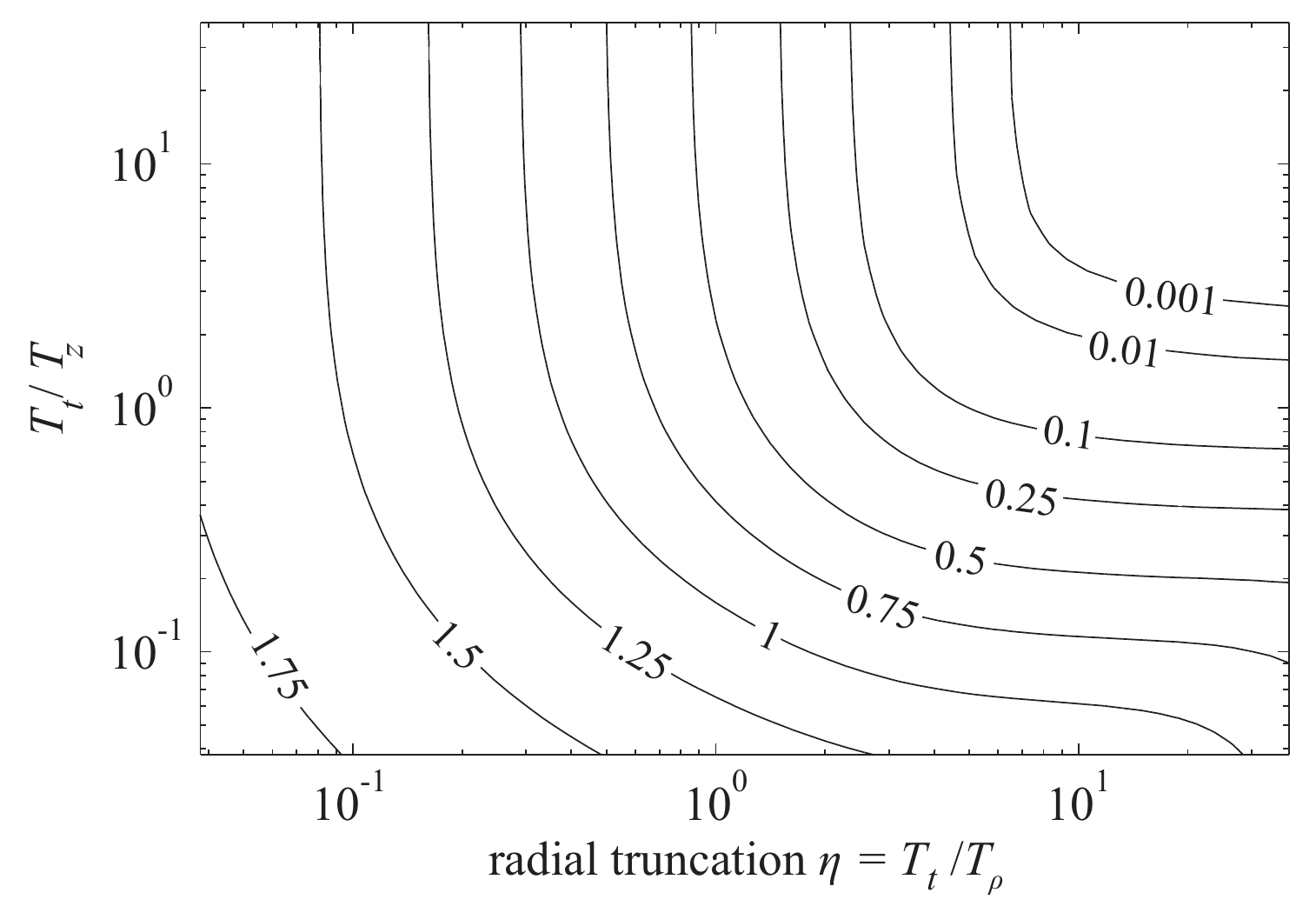}
	\caption{The function $f_{\text{loss}}(T_{\rho}, T_{z})$ defined in Eq. \eqref{eq:f_loss_t} is plotted with respect to generalized coordinates. The x-axis corresponds to the truncation parameter. Similarly, the y-axis relates the radial trap depth to the axial temperature.}
  \label{fig:f_loss}
\end{figure}	
The obtained $f_{\text{loss}}(T_{\rho}, T_{z})$ is plotted in Fig. \ref{fig:f_loss}. One can see that (for high temperature parameters) on average more than one particle can be lost after a binary collision. The averaged loss or gain of kinetic energy after one collision can now be written as the difference between the mean thermal energy before $\bar{\epsilon}$ and after a collision $\bar{\epsilon}^{+}$:
\begin{align}
	\Delta \bar{\epsilon}_{\rho} \, =& \, \bar{\epsilon}^{+}_{\rho}\, - \, \bar{\epsilon}_{\rho} \label{eq:eps1}, \\
	\Delta \bar{\epsilon}_{z} \, =& \, \bar{\epsilon}^{+}_{z}\, - \, \bar{\epsilon}_{z} \label{eq:eps2}.
\end{align}
The considerations so far and in particular Eqns. \eqref{eq:eps1} and \eqref{eq:eps2} only include kinetic energy, i.e. they apply to a trap shape where potential energy can be neglected, as in a box of finite size and infinitely steep walls. In a harmonic potential, however, a particle on time average has an equal amount of kinetic and potential energy $\bar{\epsilon}_{\text{pot}}=\bar{\epsilon}_{\text{kin}}$. During an impact, the kinetic energy ${\epsilon}_{\text{kin}}=\vec{p}^{2}/2m$ is redistributed among the colliding particles as described by the model derived above, while both colliding atoms retain their fraction of potential energy before the collision. To account for the potential shape, we assume in the following that only half of a particle's total energy can be altered in a single collision event:
\begin{align}
	\Delta \bar{\epsilon}_{\rho}  =& \frac{1}{2}\bar{\epsilon}^{+}_{\rho} + \frac{1}{2}\left(2-f_{\text{loss}}\right)\bar{\epsilon}_{\rho} -\bar{\epsilon}_{\rho} = \frac{1}{2}\left(\bar{\epsilon}^{+}_{\rho}-f_{\text{loss}} \, \bar{\epsilon}_{\rho} \right) \\
	\Delta \bar{\epsilon}_{z} =& \frac{1}{2}\bar{\epsilon}^{+}_{z} + \frac{1}{2}\left(2-f_{\text{loss}}\right)\bar{\epsilon}_{z} -\bar{\epsilon}_{z} = \frac{1}{2}\left(\bar{\epsilon}^{+}_{z}-f_{\text{loss}} \, \bar{\epsilon}_{z} \right) .
\end{align}
The shape of the potential determines the distribution between kinetic and potential energy \cite{Clausius1870} and in this way it is responsible for the speed of cross-dimensional thermalization.
\subsection{Collision rate}
\label{sec:collrate}
In Sec. \ref{sec:thermodynamics} we derived the statistical outcome of one single collision. In the following we assume that the evolution of the trap (considering $N$, $E_{\rho}$, and $E_{z}$) is determined by the product of the collision rate and the mean effect of a single collision \footnote{I.e.: $\int f_{\text{loss}}(p_{\rho},q) \cdot \sigma(q) \cdot f_{q}(q) \:  dq  \: \approx \: \int \sigma(q) \cdot f_{q}(q) \:  dq \: \cdot \: \int f_{\text{loss}}(p_{\rho},q) \cdot f_{q}(q) \:  dq $. Since the collision rate is a function of the relative momentum between colliding particles, like the statistical result of one collision $f_{\text{loss}}$, $\bar{\epsilon}_{\rho}^{+}$ and $\bar{\epsilon}_{z}^{+}$ (defined in Eqn. \eqref{eq:f_loss} - \eqref{eq:eps_z}), this assumption strictly is only true in thermal equilibrium.}. The rate at which these events occur is given by the elastic collision rate $\Gamma_{\text{coll}}$ which scales with the product of average density, scattering cross section $\sigma$, and the mean relative velocity. \par
To obtain the trap's effective volume for evaporation, the phase space distribution of a classical ideal gas at thermal equilibrium \cite{walraven96}
\begin{align}
	f\left(\textbf{r}, \textbf{p}\right) = n_{0} \left(\frac{2 \pi \hbar}{\pi m k_{B}T}\right)  e^{-\left[U(\textbf{r})+p^{2}/2m\right]/k_{B}T}
	\label{eq:f}
\end{align}
is integrated over all momentum states that occur in the trap potential: $n(\textbf{r})=(2 \pi \hbar)^{-1} \int d^{3}p f(\textbf{r},\textbf{p})$.
With the discussed properties of the potential (separability and non-uniformity) this leads to:
\begin{align}
	n(\textbf{r}) = n_{0}\left(1 - e^{\left[\epsilon_{\rho} - U_{\rho}(\rho)\right]/k_{B}T_{\rho} } \right) e^{-U_{\rho}/k_{B}T_{\rho}} \cdot e^{-U_{z}/k_{B}T_{z}}.
\label{eq:n}
\end{align}
The effective trap volume ${V}_{e}$ of the trapped gas for evaporation is defined by \cite{Yang07}:
\begin{align}
	V_{e}\equiv \frac{N^{2}}{\int n^{2}(\vec{r}) \: d^{3}r} \: .  \label{eq:Ve}
\end{align}
For $\sigma$, we use the energy dependent s-wave scattering cross section in the effective range approximation \cite{Schmidt:2003.1}:
\begin{align}
		\sigma(k) \: = \: \frac{\: 8 \pi a^{2}}{k^{2} a^{2} \: +\:\left(1/2 \: k^{2} \: a\: r_{e} - 1 \right)^{2} } \: .
\end{align}
with $a = 103 a_{0}$ and the effective range $r_{e} = 74 a_{0}$ \cite{Werner05}. $a_{0}$ denotes the Bohr radius. Contributions of higher order partial waves are neglected although they are likely to play a role at the present temperature range. However, related experimental or theoretical data are lacking \footnote{For chromium, $\sigma$ is known in the purely s-wave temperature range below $10^{-4}$ K \cite{Stuhler:2001.1} and has been measured above $10^{-2}$ K \cite{Weinstein:2002.1}. For the window in between, however, neither theoretical nor experimental data are available.}.\par
We define the collision parameter $\gamma_{\text{coll}}$ through the relation \cite{Yang07}:
\begin{align}
	\Gamma_{\text{coll}}=\gamma_{\text{coll}} \: N^{2} .
\end{align}
This leads to the definition of the thermally averaged collision parameter:
\begin{align}
	\gamma_{\text{coll}}(T_{\rho},T_{z}) \: = \: \frac{\int_{0}^{\infty} 2 \,q \: \sigma (2\,q)  \: f_{q}(q,T_{\rho},T_{z}) \: dq}{m \: V_{e}(T_{\rho},T_{z}) }.
	\label{eq:gcoll}
\end{align}
The relative wave vector $k$ in the scattering cross section is substituted by the relative momentum coordinate $q=\hbar k$, where the factor of 2 in front of $q$ stems from the relative mass $\mu=m/2$.
\subsection{The rate equations}
Together, the considerations from the previous paragraphs result in the following set of coupled differential equations:
\begin{widetext}
\begin{align}
	\dot{N} \: &= \: L \: -\gamma N \: - \: \gamma_{\text{coll}}(T_{r},T_{z})  \: f_{\text{loss}}(T_{r},T_{z})  N^{2} \label{eq:rate_n} \\
	\dot{E_{\rho}} \: &= \: \bar{\epsilon}_{L, \rho} L \:- \:\gamma \bar{\epsilon}_{\rho} N \: - \: \gamma_{\text{coll}}(T_{\rho},T_{z}) \cdot \left[\bar{\epsilon}_{\rho}^{+}(T_{r},T_{z})  - f_{\text{loss}}(T_{r},T_{z})  \: \bar{\epsilon}_{\rho}\right] N^{2} \label{eq:rate_er} \\
	\dot{E_{z}} \: &= \: \bar{\epsilon}_{L, z} L \:- \: \gamma \bar{\epsilon}_{z} N \: - \: \gamma_{\text{coll}}(T_{\rho},T_{z}) \cdot \left[\bar{\epsilon}_{z}^{+}(T_{r},T_{z})  - f_{\text{loss}}(T_{r},T_{z})  \: \bar{\epsilon}_{z} \right] N^{2} \label{eq:rate_ez}
\end{align}
\end{widetext}
with
\begin{align}
	T_{z} &=\frac{2 \epsilon_{z}}{k_{B}}=\frac{2 E_{z}}{N k_{B}} \\
	T_{\rho} &= T_{\rho}(\epsilon_{\rho})=T_{\rho}(\frac{E_{\rho}}{N}) \label{eq:e2T}
\end{align}
where Eq. \eqref{eq:e2T} is the inverse function of Eq. \eqref{eq:T2e}. $E_{\rho}$ and $E_{z}$ are the total kinetic energies in the radial and axial degrees of freedom, respectively. The first terms on the right-hand side of Eqn. \eqref{eq:rate_n}-\eqref{eq:rate_ez} describe the loading process. Particles are inserted into the trap at a loading rate $L$, a mean kinetic energy of $\bar{\epsilon}_{L, z}$ axially and $\bar{\epsilon}_{L, \rho}$ radially.
Single body loss processes such as background gas collisions and fast atoms from the beam are covered by the second terms with the respective coefficient $\gamma$. Finally, the last terms account for binary elastic collisions among trapped atoms. They include plain evaporation and cross-dimensional thermalization within the model developed in this section.
\section{The experiment}
\subsection{The loading process}
\label{sec:load}
\begin{figure*}
	\centering
				\includegraphics[width=0.99\textwidth]{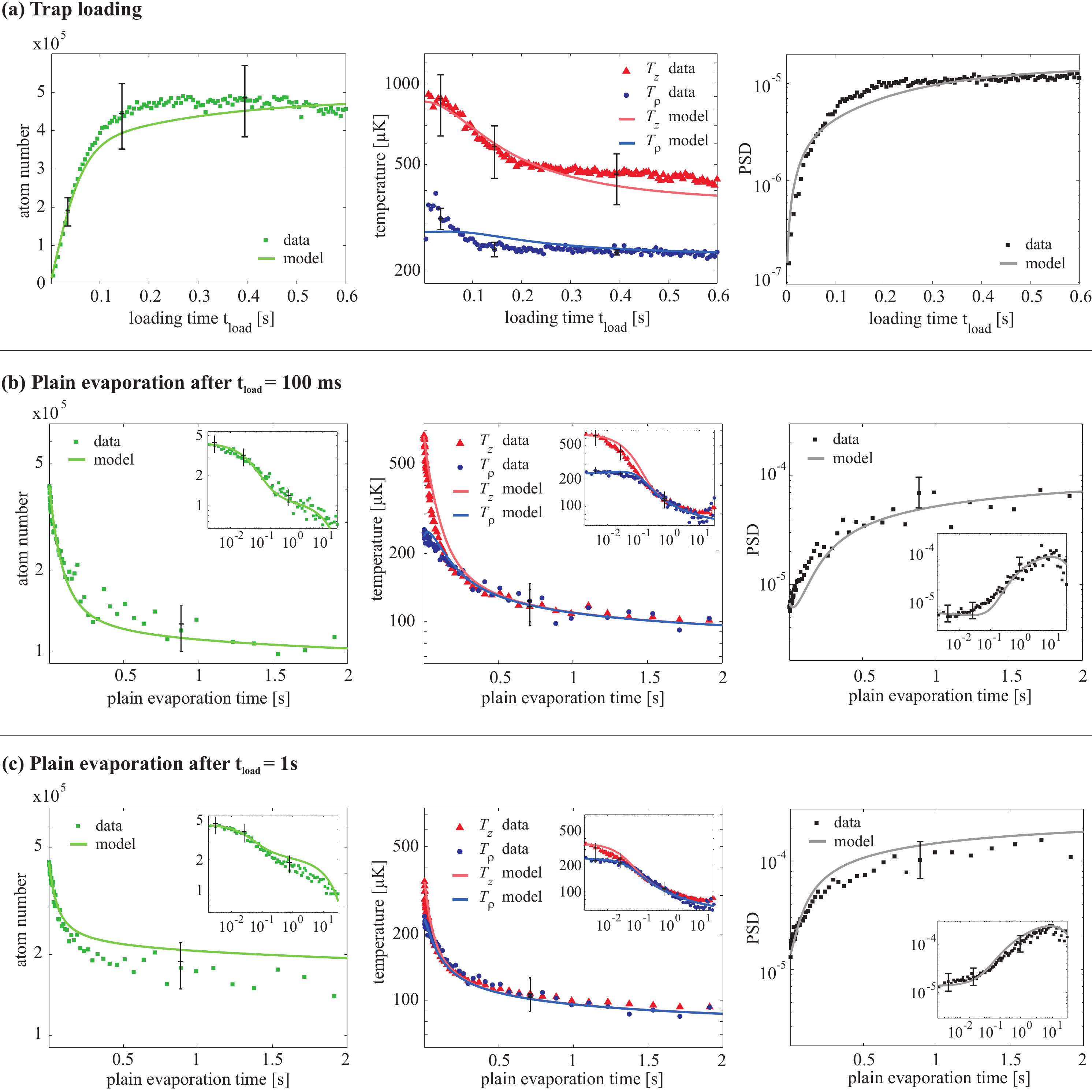}	
	\caption{(Color online) Evolution of the number of atoms $N$, the temperature parameters ($T_{\rho}$ and $T_{z}$) and the PSD during the loading process \textbf{(a)} and successive plain evaporation after $t_{\text{load}}=100\:$ms \textbf{(b)} and $t_{\text{load}}=1\:$s \textbf{(c)}. Data are extracted from in-trap absorption images (squares, triangles, and circles), while the solid lines show computations according to the model developed in Sec. \ref{sec:loading_rate_eq}. Displayed errorbars result from an estimated uncertainty in the magnification of the imaging system of 10\%. }
	\label{fig:ladekurve}
\end{figure*}
Fig. \ref{fig:ladekurve}(a) shows the evolution of the atom number, temperature, and PSD during the loading process. At time ${t}=0$, the optical pumping beam is switched on and thereby the loading process starts. Experimental data extracted from in-trap absorption images (squares, triangles, and circles) are compared to the model developed in Sec. \ref{sec:loading_rate_eq}. The theoretical curves (solid lines) are based on the numerical integration of Eqn. \eqref{eq:rate_n}. The set of coupled differential equations require three starting values: $N_{0}$, $E_{\rho,0}$, and $E_{z,0}$. Since the loading process starts from an empty trap, these parameters are all zero. Furthermore, the equations contain the parameters $\gamma$, $\bar{\epsilon}_{L, \rho}$, $\bar{\epsilon}_{L, z}$, and $L$. $\gamma$ is estimated from decay measurements of the fully loaded trap in presence of the atomic beam. It lies well above \unit[1]{s} and therefore does not play a role for the loading process under typical experimental conditions. The remaining parameters are directly deduced from the respective experimental data, within their respective experimental uncertainty. $L$ is obtained from a slope of $N(t)$ in the initial stage. $\bar{\epsilon}_{L, \rho}$ and $\bar{\epsilon}_{L, \rho}$ correspond to the temperature parameters $T_{\rho}$ and $T_{z}$ (within their experimental uncertainties) of the cloud before collisions play a role.\par
One can see that during the first \unit[50]{ms} of loading, the atom number increases at a constant rate. A linear fit yields a loading rate of \unit[$5.8 \cdot 10^{6}$]{atoms/s}, which corresponds to a loading efficiency of \unit[10-15]{\%} with respect to the flux in the atomic beam. After about \unit[100]{ms} $N$ saturates around $N = 5\cdot 10^{5}\:$atoms. As also shown in Fig. \ref{fig:ladekurve}(a), the temperature decreases during the loading process.
\subsection{Plain evaporation}
As a benchmark of the loading scheme it is interesting to validate the agreement between the model and experimental data under different conditions, i.e. after the loading process is terminated and thus $L=0$. In Eqn. \eqref{eq:rate_n}-\eqref{eq:rate_ez} the first terms on the right hand side vanish. $\gamma$ is smaller than during the loading process. Since the atomic beam is switched off, the $1/e$ lifetime is now solely limited by background pressure to \unit[30]{s}. The remaining starting conditions for the simulation are: $N_{0}$, $E_{\rho,0}$, and $E_{z,0}$. These are obtained from measurements that are taken directly after the loading process has been terminated. \par
Figs. \ref{fig:ladekurve}(b) and (c) show the temporal evolution of the trap directly after the loading process has been stopped with a preceding ${t}_{\text{load}}=$\unit[100]{ms} (b) and ${t}_{\text{load}}=$\unit[1]{s} (c), respectively. Data collected during plain evaporation are compared to the model developed in Sec. \ref{sec:loading_rate_eq}. The plots in the top row show the evolution of the atom number, while below the respective temperature measurement is displayed. The plots cover the first two seconds, which is the experimentally most relevant time frame. To emphasize the dynamics at small and large time scales, the insets have an extended x-axis up to \unit[30]{s} and a double logarithmic scale. \par
From the evolution of $N$ in Figs. \ref{fig:ladekurve}(b) and (c), one can see a fast decay during the first \unit[500]{ms} that can be attributed to plain evaporation before eventually, the decay becomes exponential. As expected, the initial fast decay is more pronounced in (b), where the trapped atoms have a higher mean energy compared to (c).
Concerning temperature, one observes $T_{\rho}$ and $T_{z}$ equilibrate within about \unit[50]{ms}, as well as a decreasing mean temperature throughout plain evaporation. Despite the atom losses, the PSD is increasing by about one order of magnitude during plain evaporation, which indicates that evaporative cooling due to elastic collisions is the dominant process.
\subsection{Limitations to the loading process}
As can be seen from Fig. \ref{fig:ladekurve}, the number of atoms saturates after a trap loading time of less than \unit[200]{ms}. It would be favorable to prolong the loading phase and reach a higher equilibrium atom number $N_{\text{max}}$. Therefore, in the following, we investigate the mechanism that is responsible for the saturation of $N$. The good agreement between experimental data and simulations suggests plain evaporation to be the dominant loss process. No other losses are included into the model, apart from single body losses, which are yet on another time scale.
\begin{figure}
	\centering
		\includegraphics[width=0.45\textwidth]{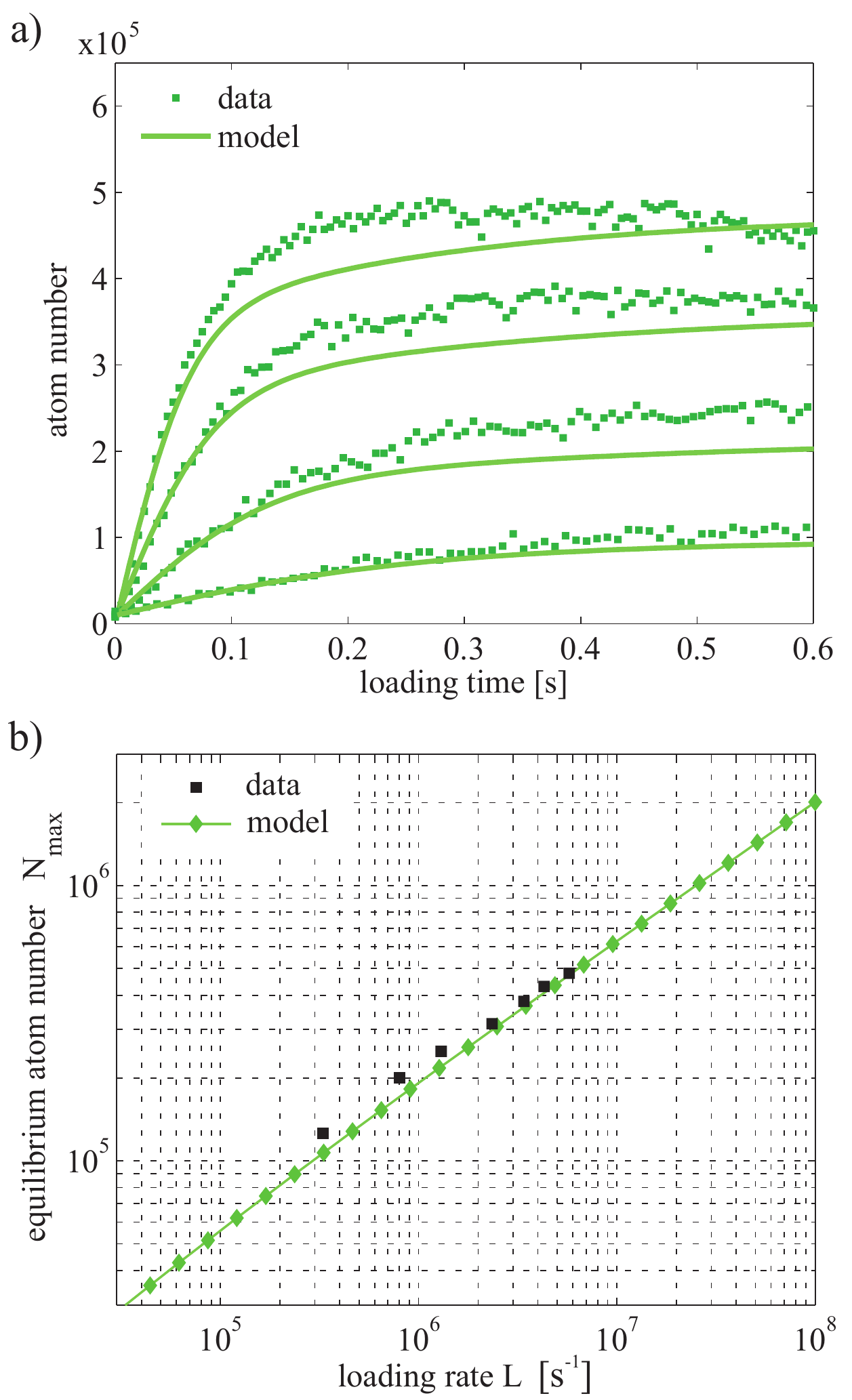}
	\caption{(Color online) (a) The temporal evolution of the atom number is plotted for data corresponding to different loading rates $L$. The squares represent experimental data and are compared to simulations (solid line), which merely differ in the loading rate $L$ that is measured directly from the initial slope of the experimental data. (b) The squares show the equilibrium atom number $N_{\text{max}}$ measured after a long $t_{\text{load}}$ with respect to the corresponding loading rate $L$. The diamonds connected by a solid line show $N_{\text{max}}$ from the simulations.}
  \label{fig:loadproc}
\end{figure}	
Fig. \ref{fig:loadproc}(a) shows simulations and experimental data of the loading process with different loading rates. We experimentally reduce $L$ by decreasing the flux of Cr atoms between our effusion cell and the MOT. The measured loading rate is then used in the respective simulations, while all other parameters remain the same.
When an elastic two-body loss process such as evaporation constitutes the dominant source of loss, the stationary solution of the simplified rate equation for the atom number Eq. \eqref{eq:ratesimple} is of the following form (i.e. $\gamma_{ev} N \gg \gamma$):
\begin{align}
	N_{\text{max}}=\sqrt{\frac{L}{\gamma_{ev}}}.
	\label{fig:2bodyloss}
\end{align}
$N_{\text{max}}$ is plotted as a function of $L$ in Fig. \ref{fig:f_loss}(b). Both simulations and experimental data agree with the behavior described by Eq.\eqref{fig:2bodyloss}. It might seem counter-intuitive that elastic collisions between trapped atoms can actually limit $N_{\text{max}}$. Atoms are loaded into the trap with a mean energy corresponding to $\eta \approx 2$. The mechanism can therefore be understood in the context of the continuous heating rate imposed by the loaded atoms and the cooling rate due to evaporation. An equilibrium temperature develops where $\gamma_{\text{coll}}(T_{r},T_{z})  \cdot f_{\text{loss}}(T_{r},T_{z}) = \gamma_{2} = \text{const}$. The key to increase $N_{\text{max}}$, is therefore a reduction of $f_{\text{loss}}$. One way to achieve this is to increase the radial trap depth, e.g. by a higher ODT power. Simultaneously increasing the ODT's waist maintains the trap frequencies and thus avoids additional radial heating. Our computations predict that loading the trap at $\eta=3$ results in gain by a factor of 2 in $N_{\text{max}}$ compared to the current experimental parameters (at equal $L$). At the same time, the steady state PSD improves by more than a factor of 7.
\subsection{Forced evaporation}
Plain evaporation as described by the model developed in Sec. \ref{sec:loading_rate_eq} quickly slows down with increasing $\eta$. To maintain evaporative cooling, the trap depth has thus to be continuously lowered (forced evaporation). In our case, we realize this by decreasing the ODT's power. There is a large range of literature on evaporative cooling in magnetic traps \cite{Pinkse98, Walraven:96, Berg97, Surkov96} or in optical dipole traps \cite{OHara01, Adams95}. The hybrid magnetic and optical trap geometry, as a prerequisite of the continuous loading process, combines advantages of either trap type. On the one hand, the high-field seeking absolute ground state can be trapped, while on the other hand, the trap weakens the typical problem of a single beam ODT: the decreasing collision rate as a result of the coupling between the trapping frequencies and trap depth. Here, the axial confinement is maintained throughout the evaporation ramp by the magnetic field, while merely the radial trapping frequency is lowered. In order to extend the model described above for forced evaporation, we have to consider the work exerted on the trapped gas as a result of the variation of the radial trap shape \cite{Pinkse98}:
\begin{align}
	E_{\text{ext}}=\frac{N k_{B}T}{V_{e}(T_{\rho}, T_{z})}\left(\frac{\partial V_{e}}{\partial P_{ODT}}\right) \frac{d P_{ODT}}{dt}.
	\label{eq:work}
\end{align}
Spilling \cite{Walraven:96}, i.e. the loss of trapped states due to changes in the potential shape can be neglected for our experimental conditions ($\eta \approx 7$ during the ramp).\par

\begin{figure}
	\centering
		\includegraphics[width=0.45\textwidth]{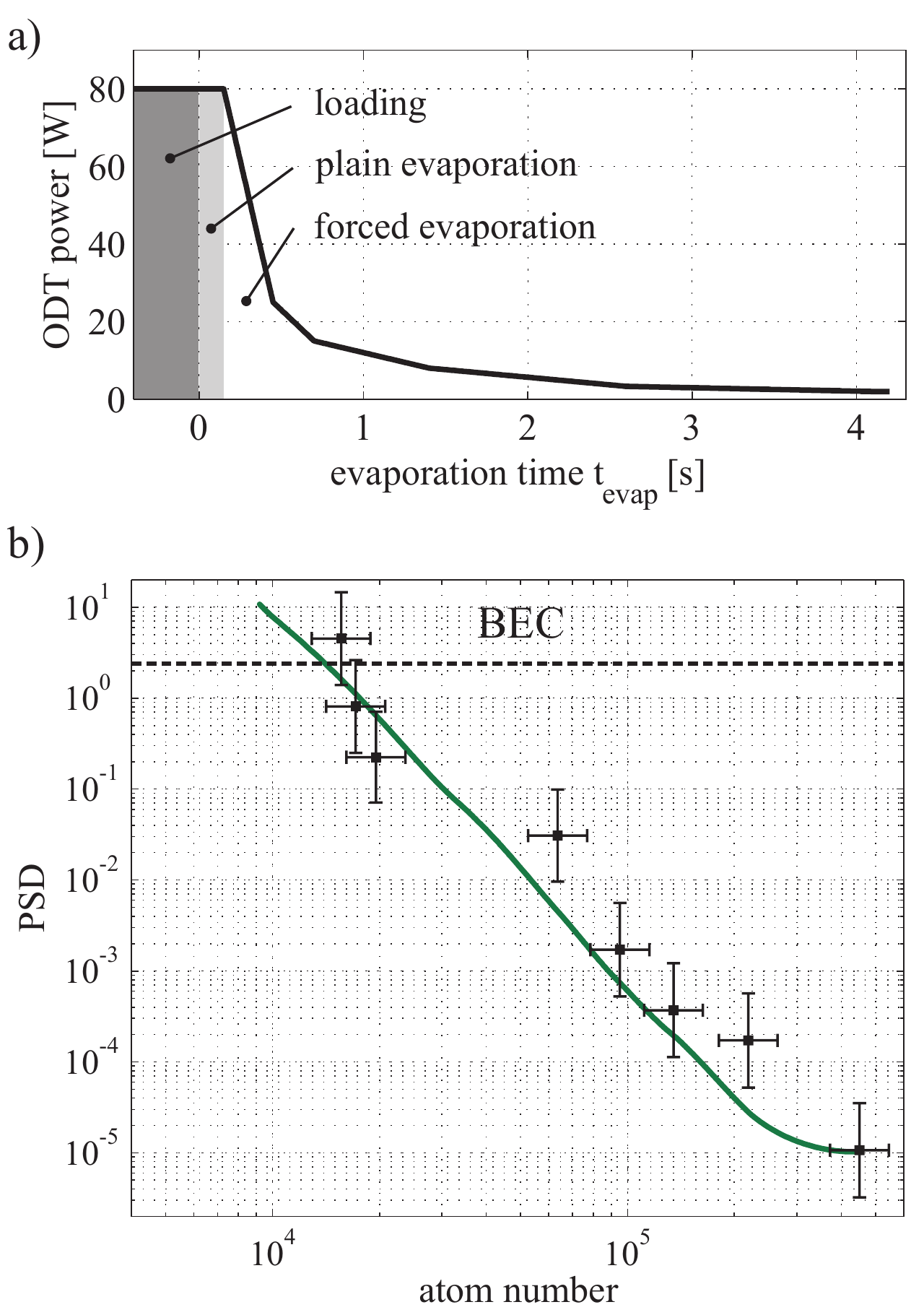}
	\caption{(Color online) (a) Experimentally determined ramp of the ODT power that is used to produce a BEC. At $t=0$ the loading process is stopped and followed by plain and forced evaporation as indicated. (b) Evolution of the PSD as a function of the atom number during evaporation. The squares are data points that are measured during the evaporation ramp. The error bars result from fit uncertainties and systematic errors as the magnification of the imaging system. The solid line represents a simulation of the forced evaporation process according to the classical model discussed in the text.}
  \label{fig:npsd}
\end{figure}	

Fig. \ref{fig:npsd}(a) shows the experimentally optimized ramp of the ODT power for forced evaporative cooling. It consists of one plain evaporation segment starting at $t=0$ and subsequently 5 segments with linearly decreasing ODT power. In (b), the evolution of the PSD is shown with respect to $N$. The squares show data taken during the evaporation ramp, while the solid line indicates the respective prediction according to model of the rate equations. It should be noted, however, that these do not consider quantum statistical effects. After forced evaporative cooling, we reach a BEC with $N \approx 10^{4}$ atoms in an overall time of \unit[4.5]{s}.
\section{Conclusion}
In conclusion, we have developed a model based on a the set of coupled rate equations which is capable of modeling continuous loading schemes for an ODT from a beam of atoms \cite{anoush09, falkenau11}. Due to the non-uniform trap geometry and the simultaneous appearance of loading and evaporation, the dynamics of the thermodynamical properties of the trapped cloud are far more complicated than in previously modeled systems. As a direct result of the implementation of the loading scheme, the dimensionality of evaporation \cite{Surkov96} is two for the non-uniform trap geometry discussed here.\\
We have shown that numerical simulations of the loading process, plain, and forced evaporation agree quantitatively well with experimental data. In experiments as well as in numerical simulations we observe the onset of saturation at an atom number of about \unit[$5\cdot 10^{5}$] after loading the trap for \unit[100]{ms}. The model allows identifying elastic binary collisions as the limiting process within our current experimental conditions. It is important to remark that the system is not yet limited by light assisted collisions mediated by (re)absorbed pump-light photons. Larger steady state densities can therefore be reached through a higher trap parameter $\eta$ of the atoms loaded into the trap. This may be realized by means of a radially colder beam or a higher radial trap depth.\\
Furthermore, we have shown that the specific trap geometry has good properties for successive evaporative cooling, since the axial trap frequency is independent from the trap depth. In combination, the loading mechanism followed by forced evaporation in the hybrid magnetic and optical trap allows producing a Cr BEC within less than \unit[5]{s}, which is significantly faster than previously published \cite{griesmaier05, beaufils08} experiments on Cr atoms.\\
The demonstrated loading mechanism may as well be applicable to atoms where a closed cycling transition for laser cooling is not available, or for molecules, which can except for a few cases not be laser cooled due to their complex level structure. Thereby, good starting conditions for successive evaporative cooling can be reached, circumventing the need for laser cooling on a closed cycling transition -- given the atomic or molecular species can be prepared in a cold and slow beam. It may also serve as a continuous source of ultra-cold matter that may constitute a step towards a continuous wave atom laser.

\end{document}